\documentclass[
  aps,prl,twocolumn,superscriptaddress,floatfix,
  amsmath,amssymb,nofootinbib,longbibliography,secnumarabic
]{revtex4-2}

\usepackage{url}
\usepackage{mathtools}
\usepackage{braket}
\usepackage{physics}
\usepackage{soul}
\usepackage[english]{babel}
\usepackage{bm}
\usepackage{comment}
\usepackage{color}
\usepackage{graphicx}
\usepackage{orcidlink}
\usepackage{bibunits}

\usepackage{hyperref}
\hypersetup{
  colorlinks = true,
  citecolor  = blue,
  linkcolor  = blue,
  urlcolor   = blue
}

\usepackage[normalem]{ulem}

\makeatletter
\let\label\ltx@label
\makeatother

\newcommand{\runinhead}[1]{\textit{#1.}\ }

\newcommand{\SigAnom}{ \Sigma^{12(\infty)} }
\newcommand{\SpVars}{\mathbf{k}}

\begin{document}

\author{F.~Marino~\orcidlink{0000-0001-7743-1982}}
\affiliation{Institut f\"{u}r Kernphysik and PRISMA++ Cluster of Excellence, Johannes Gutenberg-Universität Mainz, 55128 Mainz, Germany}

\author{W. G. Jiang~\orcidlink{0000-0001-8441-972X}}
\affiliation{Institut f\"{u}r Kernphysik and PRISMA++ Cluster of Excellence, Johannes Gutenberg-Universität Mainz, 55128 Mainz, Germany}

\author{C. Barbieri~\orcidlink{0000-0001-8658-6927}}  
\affiliation{Dipartimento di Fisica ``Aldo Pontremoli'', Universit\`a degli Studi di Milano, 20133 Milano, Italy}
\affiliation{INFN,  Sezione di Milano, 20133 Milano, Italy}

\author{G.~Col\`{o}~\orcidlink{0000-0003-0819-1633} } 
\affiliation{Dipartimento di Fisica ``Aldo Pontremoli'', Universit\`a degli Studi di Milano, 20133 Milano, Italy}
\affiliation{INFN,  Sezione di Milano, 20133 Milano, Italy}

\title{\textit{Ab initio} Green's function theory of superfluid neutron matter}


\begin{abstract}
At the low densities encountered in the inner crust of neutron stars, neutron matter becomes superfluid, directly influencing macroscopic phenomena such as star cooling.
However, a precise quantitative understanding of this system has been hindered by uncontrolled many-body approximations.
In this letter, we exploit the nonperturbative Gorkov algebraic diagrammatic construction method to provide a unified \textit{ab initio} description of the equation of state and the single-particle spectral functions of homogeneous neutron matter, spanning from the dilute regime to saturation density.
We identify clear microscopic signatures of superfluidity in the excitation spectrum and quantify the pairing gaps associated with realistic chiral Hamiltonians, estimating for the first time their sensitivity to the many-body truncation.
Our robust handling of both dynamical and pairing correlations paves the way for exploring electroweak processes across the regimes relevant to neutron stars.
\end{abstract}

\maketitle

\runinhead{Introduction}
Homogeneous nuclear matter, an extended system of nucleons interacting through the strong force, bridges the vastly different scales of finite nuclei and neutron stars~\cite{Gandolfi2015,Oertel2017,Burgio2021,Chatziioannou2024}.
A combination of laboratory experiments~\cite{Mammei:2023kdf,vonNeumann-Cosel:2025hsj}, astrophysical observations~\cite{cromartie2020relativistic,Raaijmakers_2021}, and accurate nuclear \textit{ab initio} computations~\cite{Lovato2022PNM,Marino2024,Hu2026}, often based on Hamiltonians derived within chiral effective field theory~\cite{Epelbaum2024,MACHLEIDT2024104117}, provides constraints on the equation of state (EOS) of neutron matter at the intermediate densities characteristic of the outer core of neutron stars ($\rho \approx 1$--$2\,\rho_0$, where $\rho_0 \approx 0.16 \,\mathrm{fm}^{-3}$).

A different physical regime characterizes the inner crust, where neutron-rich nuclei coexist with a dilute neutron gas at nearly zero temperature~\cite{Chamel2008Crust,HaskellSedrakian2018}.
At such low densities, neutron-neutron interactions are dominated by the attractive $^1S_0$ channel, rendering the system unstable to the formation of Cooper pairs and driving the onset of superfluidity~\cite{Dean2003, Sedrakian2006,Gezerlis2014Review,Sedrakian2019,Ramanan:2020xrj}. Superfluidity modifies neutrino emission and absorption processes~\cite{Sedrakian2019,Yakovlev2001,Sedrakian2007,Steiner2009} and the rotational dynamics (including pulsar glitches) of neutron stars~\cite{HaskellSedrakian2018,Skao}.
Vice versa, observations of the cooling of isolated neutron stars might help constrain microscopic properties, for example, the magnitude of singlet gaps~\cite{WeiBurgio2020} or the critical temperature for triplet pairing~\cite{Page_2009,Beloin2019}.

The standard description of nuclear superfluidity is based on BCS mean-field theory, which predicts pairing gaps, the energy required to break a Cooper pair, under the assumption of weak interactions~\cite{Gezerlis2014Review,Sedrakian2019}. This assumption is however questionable in strongly correlated nuclear systems and has motivated beyond-mean-field computations (see e.g.~\cite{Cao2006, Gezerlis2010,Wlazlowski2011,Gezerlis2014Review, Ding2016pairing,Drissi2022, Gandolfi2022, DataPy, Urban2026, Multimodal}).
Although beyond-mean-field corrections are generally found to suppress the $^1S_0$ gap relative to BCS predictions~\cite{Burgio2021}, superfluid pairing gaps remain substantially less constrained than the EOS.
This owes to both their exponential sensitivity to medium effects~\cite{Sedrakian2019} and the lack, in past calculations, of an assessment of theoretical uncertainties.
The use of uncontrolled approximations, in particular, has hindered, so far, a rigorous understanding of the pairing properties of chiral Hamiltonians beyond the mean-field theory~\cite{Drischler2017,Drissi2022}.

\begin{figure}[h!]
    \centering
    \includegraphics[width=\columnwidth]{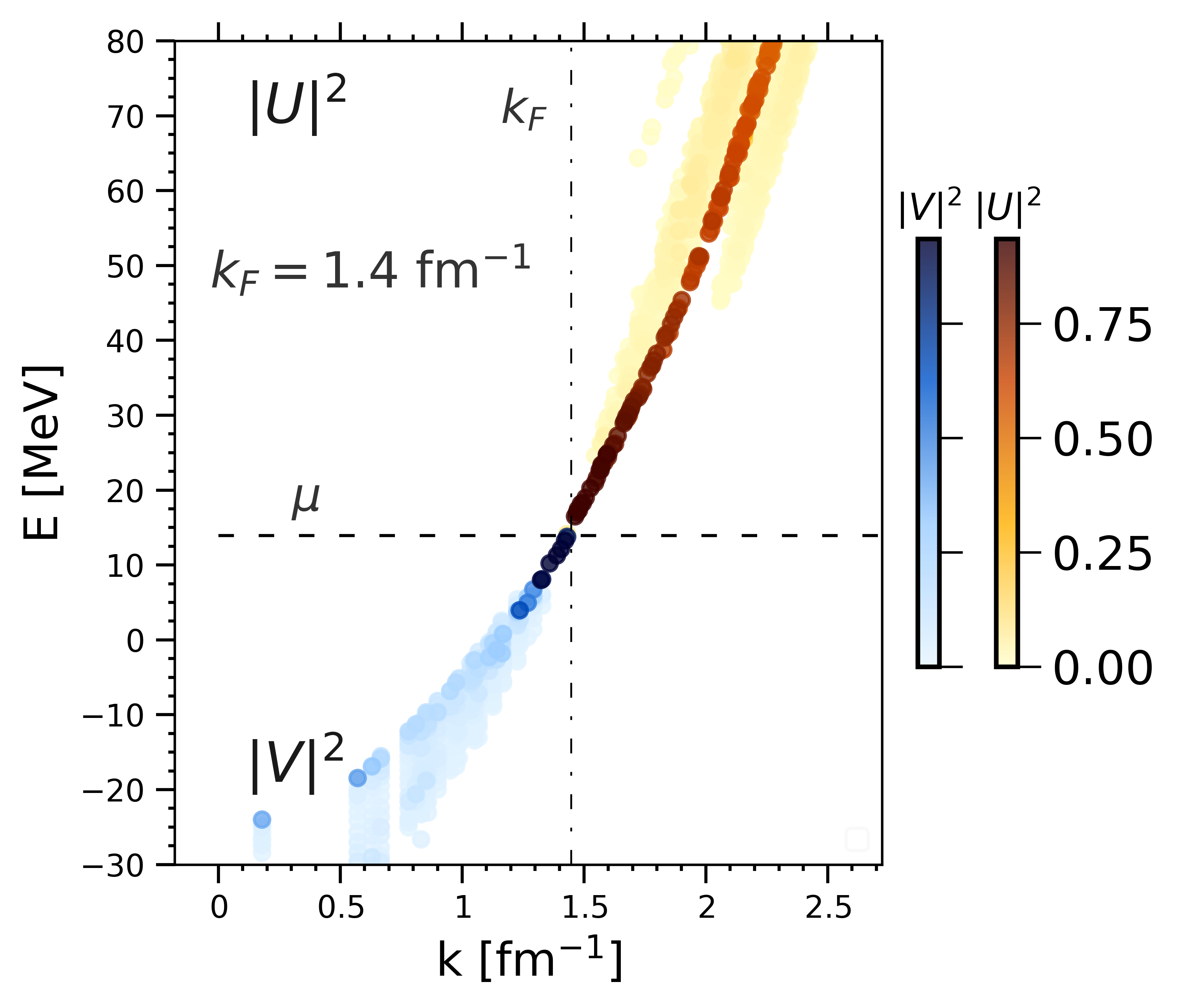}
    \includegraphics[width=\columnwidth]{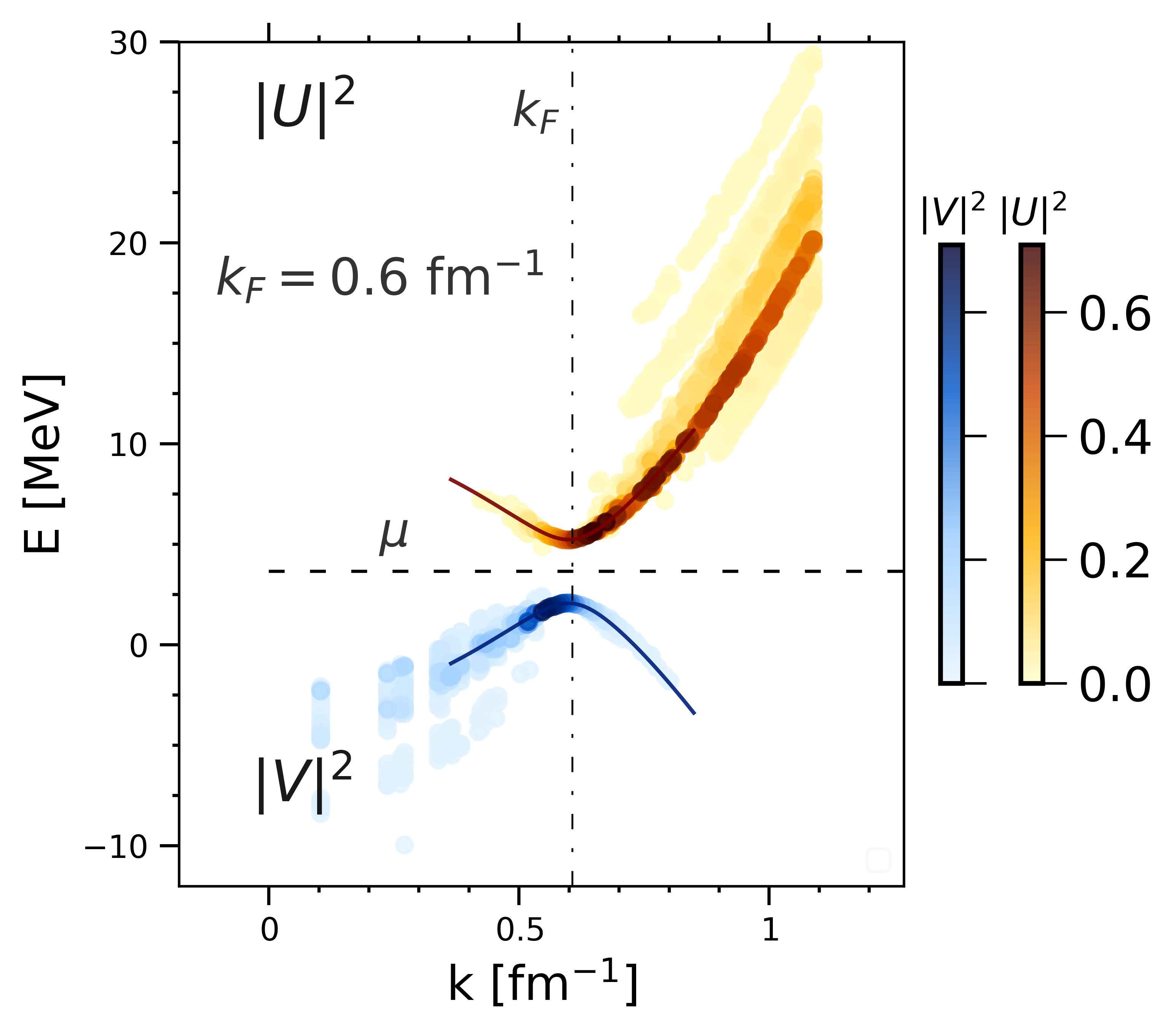}
    \caption{
    Two-dimensional representation of the spectral function~\eqref{eq: spectral function 11}
    obtained by varying the momentum $k$ and energy $E=\mu+\hbar\omega$.
    The top panel displays neutron matter in the normal-liquid regime (at Fermi momentum $k_F=1.4\, \mathrm{fm}^{-1}$), while in the bottom panel ($k_F=0.6 \, \mathrm{fm}^{-1}$) the system is in the superfluid phase.
    The squared spectroscopic amplitudes $\abs{\mathcal{U}^{q}_{\mathbf{k}}}^2$ and $\abs{\mathcal{V}^{q}_{\mathbf{k}}}^2$ are shown for poles above and below the Fermi level $E=\mu$, respectively. The corresponding color scales are displayed alongside each panel. The Fermi momentum $k_F$ and chemical potential $\mu$ are indicated by dot-dashed and dashed lines, respectively. For $k_F=0.6\,\mathrm{fm}^{-1}$, continuous lines show fits of the spectral function obtained with Eq.~\eqref{eq: dispersion} (see text). 
    Results are obtained using ADC(3) with sp-TABCs and the $\rm{NNLO_{sat}}(450)$ interaction~\cite{NNLOsat}.
    }
    \label{fig:spectral functions}
\end{figure}

In this letter, we determine the pairing gaps from full-fledged simulations of the Landau quasiparticle structure of neutron matter.
We estimate, for the first time, the uncertainties associated with the many-body truncation.
These achievements are made possible by combining the nonperturbative algebraic diagrammatic construction (ADC) method with the Gorkov Green's function framework~\cite{Marino2026Gorkov,Marino2024}.
By providing momentum- and energy-resolved spectral functions, we elucidate in Fig.~\ref{fig:spectral functions} the normal-liquid and superfluid regimes of neutron matter and identify microscopic signatures of pairing correlations.

\runinhead{Methodology}
The Gorkov Self-consistent Green's function (SCGF) formalism~\cite{Barbieri2003DRPA,Soma2011,Soma2020,Barbieri2022Gorkov, Marino2026Gorkov} is intimately related to the one-body spectral function $S^{11}(\mathbf{k},\omega)$.
We consider the grand canonical Hamiltonian at zero temperature, $\Omega=H-\mu N$, with $\mu$ the chemical potential. The many-particle ground state and the excited states are denoted by $\ket{\Psi_0}$ and $\ket{\Psi_q}$, respectively, with the corresponding energies being $E_0$ and $E_q$.
The spectral function is then defined as
\begin{align}
\label{eq: spectral function 11}
S^{11}(\mathbf{k}, \omega) = \sum_q \left[ \abs{\mathcal{V}^{q}_{\mathbf{k}}}^2 \delta(\hbar\omega + \hbar\omega_q) + \abs{\mathcal{U}^{q}_{\mathbf{k}}}^2 \delta(\hbar\omega - \hbar\omega_q) \right],
\end{align}
where we have introduced the quasiparticle energies, $\hbar\omega_q = E_q - E_0$, and the amplitudes for the addition and removal of a neutron with momentum $\mathbf{k}$ as \hbox{$\mathcal{U}_{\SpVars}^{q} = \mel{\Psi_0}{c_{\SpVars}}{\Psi_q}$} and \hbox{$\mathcal{V}_{\SpVars}^{q} = \mel{\Psi_0}{c_{\SpVars}^{\dagger}}{\Psi_q}$}, respectively.
$\hbar\omega_q$ and the spectroscopic amplitudes are determined as the solution to the self-consistent Gorkov eigenvalue problem
\begin{align}
    \label{eq: Gorkov energy dependent}
    & 
    \begin{pmatrix}
        \xi_{\bf{k}} + \widetilde{\Sigma}^{11}(\mathbf{k},\omega) & \SigAnom(\mathbf{k}) \\
        \SigAnom(\mathbf{k}) & - \xi_{\bf{k}} - \widetilde{\Sigma}^{11}(\mathbf{k},-\omega)
    \end{pmatrix}
    \eval_{\omega = \omega_q}
    \begin{pmatrix}
        \mathcal{U}^{q}_{\mathbf{k}} \\ \mathcal{V}^{q}_{\mathbf{k}}
    \end{pmatrix} 
    \nonumber \\
    & = \hbar \omega_q
    \begin{pmatrix}
        \mathcal{U}^{q}_{\mathbf{k}} \\ \mathcal{V}^{q}_{\mathbf{k}}
    \end{pmatrix} .
\end{align}
Here, $\xi_{\mathbf{k}} = (\hbar \mathbf{k})^2/(2m) - \mu + \Sigma^{11(\infty)}(\mathbf{k})$.
In this setup, the Hartree--Fock--Bogoliubov (HFB) approximation\footnote{In uniform matter, HFB is equivalent to BCS, provided the normal and pairing fields are both computed self-consistently~\cite{Chamel2008}.} corresponds to including only the static normal self-energy $\Sigma^{11(\infty)}(\mathbf{k})$ and anomalous self-energy $\SigAnom(\mathbf{k})$.
Beyond-mean-field theories are generated by incorporating a dynamical self-energy $\widetilde{\Sigma}^{11}(\mathbf{k},\omega)$. 
We employ the truncation scheme of Refs.~\cite{Marino2025Qnp,Marino2026Gorkov,Marino2024}, in which $\widetilde{\Sigma}^{11}$ is computed within the ADC hierarchy~\cite{Barbieri2017,Raimondi2018ADCw3nf,Schirmer2018} and accounts for the dominant dynamical correlations in the normal channel.
We use the state-of-the-art third-order ADC approximation [ADC(3)] when not stated otherwise. 
For each Fermi momentum $k_F = (3\pi^2 \rho)^{1/3}$, neutron matter is simulated with $N=66$ particles in a cubic box of length $L$ using special-point twisted-angle boundary conditions (sp-TABC)~\cite{Marino2026Gorkov,CeperleyTABC}.
Single-particle momenta are given by $\mathbf{k} = \frac{1}{L}(2\pi \mathbf{n} + \boldsymbol{\theta})$, where $\boldsymbol{\theta}$ denotes the three twist angles and $\mathbf{n}=(n_x,n_y,n_z)\in\mathbb{Z}^3$.
While standard periodic boundary conditions (PBCs, $\boldsymbol{\theta}=0$) are very close to \hbox{sp-TABCs} for total energies, twisted angles are indispensable for the calculation of momentum-resolved observables~\cite{Marino2026Gorkov}, as the momentum grid associated with PBCs would be too coarse to reliably resolve quantities such as the pairing gap.

\runinhead{Results}
Figure~\ref{fig:spectral functions} displays our results for the spectral function of neutron matter as a function of the momentum $k$ and energy $E=\mu + \hbar\omega$, near saturation density (for Fermi momentum $k_F=1.4\,\mathrm{fm}^{-1}$) and in the superfluid regime ($k_F=0.6\,\mathrm{fm}^{-1}$).
At each momentum, Landau quasiparticle excitations are identified by the dominant peaks of the spectral function in the two half-planes $E<\mu$ and $E>\mu$.
We denote the corresponding energies by $E^-$ and $E^+$, respectively.

In the top panel, the system is in the normal phase.
The quasiparticle bands $E^{\pm}$ both follow an approximately quadratic trend as a function of $k$ and they can be continuously prolonged into each other. 
In particular, they cross the chemical potential close to $k=k_F$
\footnote{In the thermodynamic limit, the crossing takes place at \hbox{$k=k_F$}. Finite-size effects in our simulations imply that the crossing point is in general close to, but not exactly at, $k_F$. }.
This implies that neutrons near the Fermi momentum can be promoted to nearby unoccupied states at no energy cost.

The behavior is strikingly different in the superfluid phase. 
Here, the dominant particle-addition and particle-removal peaks no longer cross near $k_F$. Instead, they repel each other, leaving a finite gap in the spectral strength around the chemical potential. 
This provides a direct signature of the onset of superfluidity~\cite{li2024observation,Haussmann2009,Biss2022}

The quasiparticle bands coexist with a fragmented background of satellite poles carrying small spectral weight.
This is a consequence of dynamical correlations beyond HFB.
However, in the vicinity of the Fermi surface, the secondary solutions fade away and the neutron-matter pairing gap $\Delta$ can be quantified accurately by interpolating the two dispersion branches with the formula~\cite{Haussmann2009,Magierski2009,li2024observation}
\begin{align}
\label{eq: dispersion}
E^{\pm}(k) = \mu \pm \sqrt{
\left(
\frac{\hbar^2}{2m^{*}_{\pm} } k^2 + U_{\pm} - \mu
\right)^{2} + \Delta^2_{\pm}
},
\end{align}
where $U_{\pm}$, $m^{*}_{\pm}$, and $\Delta_{\pm}$ are the parameters of the fit (with $\Delta_{\pm}>0$).
We fit the computed spectral functions with Eq.~\eqref{eq: dispersion} and evaluate the pairing gaps as $\Delta = \frac{1}{2} (\Delta_{+} + \Delta_{-})$.
The quality of the fits can be judged from the example in the bottom panel of Fig.~\ref{fig:spectral functions}.

\begin{figure}[t]
    \centering
    \includegraphics[width=\columnwidth]{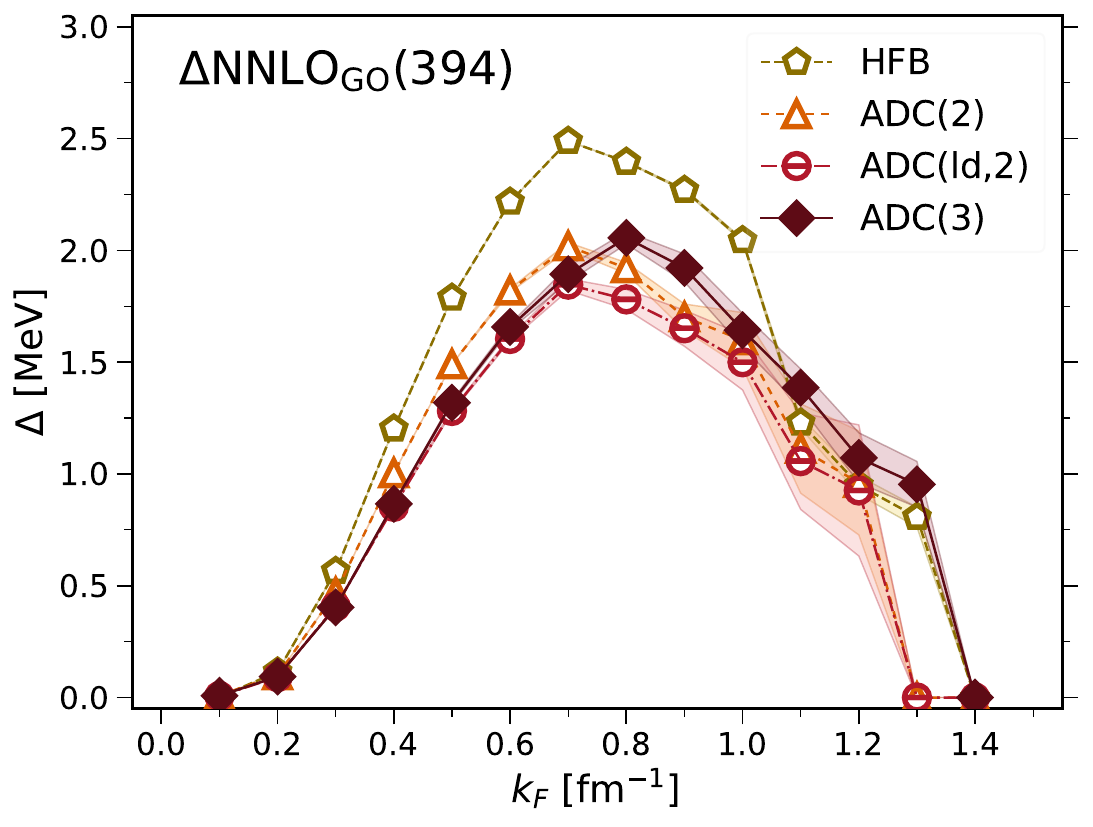}
    \caption{ Pairing gaps as functions of the Fermi momentum predicted at different approximation levels.
    Fit errors are reported as shaded bands.
    The $\Delta \rm{NNLO_{GO}}(394)$ potential from Ref.~\cite{DeltaGo2020} is used. See text.
    }
    \label{fig: Gaps_Truncations Adc}
\end{figure}

In Fig.~\ref{fig: Gaps_Truncations Adc}, we gauge the sensitivity of the pairing gaps to the many-body description by comparing predictions at different truncations in the ADC hierarchy.
Besides the baseline HFB results and ADC(3), we consider the second-order approximation [ADC(2)] and the ADC(ld,2) truncation, equivalent to an all-order resummation of ladder diagrams~\cite{Barbieri2017,Marino2024}.
The shaded bands report the errors on $\Delta$ propagated from the fit via Eq.~\eqref{eq: dispersion}.

In comparison to HFB, ADC computations tend to reduce the maximum pairing gap and shift it to slightly higher momenta.
The ADC results point to a maximum $\Delta$ of about 2 MeV for this interaction, in contrast to 2.5 MeV for HFB.

We notice a close overlap between ADC(ld,2) and ADC(3) at low Fermi momenta (up to $k_F = 0.7\,\rm{fm}^{-1}$).
The ADC(3) self-energy includes, among other contributions, the classes of both ladder and ring diagrams~\cite{Barbieri2022Gorkov,Marino2026Gorkov}. 
The impact of ladder terms, which are absent in ADC(2), is small but noticeable also at low densities.
Ring diagrams are associated with the screening of the interaction in the medium~\cite{Cao2006,Ramanan:2020xrj}, but their effects appear to be relatively modest in this region of $k_F$.

We can interpret the quenching of the gap in beyond-HFB theories by a qualitative argument.
As discussed in the Supplemental material, we can perform an expansion of the Gorkov equation around the Fermi surface ($k=k_F$, $\omega=0$) at leading order in the eigenenergies $\omega_q$ ~\cite{Lombardo2001,Idini2012}.
While in HFB the pairing gap is directly identified with the anomalous field $\SigAnom(k \to k_F)$, for  beyond-mean-field theories, the physical gap is renormalized with respect to~the HFB expression~\cite{Hebeler2009,Lombardo2001},
\begin{align}
    \Delta = \left( 1 - \frac{1}{\hbar} \pdv{\widetilde{\Sigma}^{11}}{\omega} \right)^{-1} \SigAnom(k_F)
\end{align}
where the derivative of the self-energy is evaluated for $k=k_F$ and $\omega=0$, and $\pdv{\widetilde{\Sigma}^{11}}{\omega} < 0$. 
The quenching factor is the energy-dependent contribution to the effective mass $m^*/m$~\cite{LombardoSchulze_notes}.

\begin{figure}[ht!]
    \centering
    \includegraphics[width=\columnwidth]{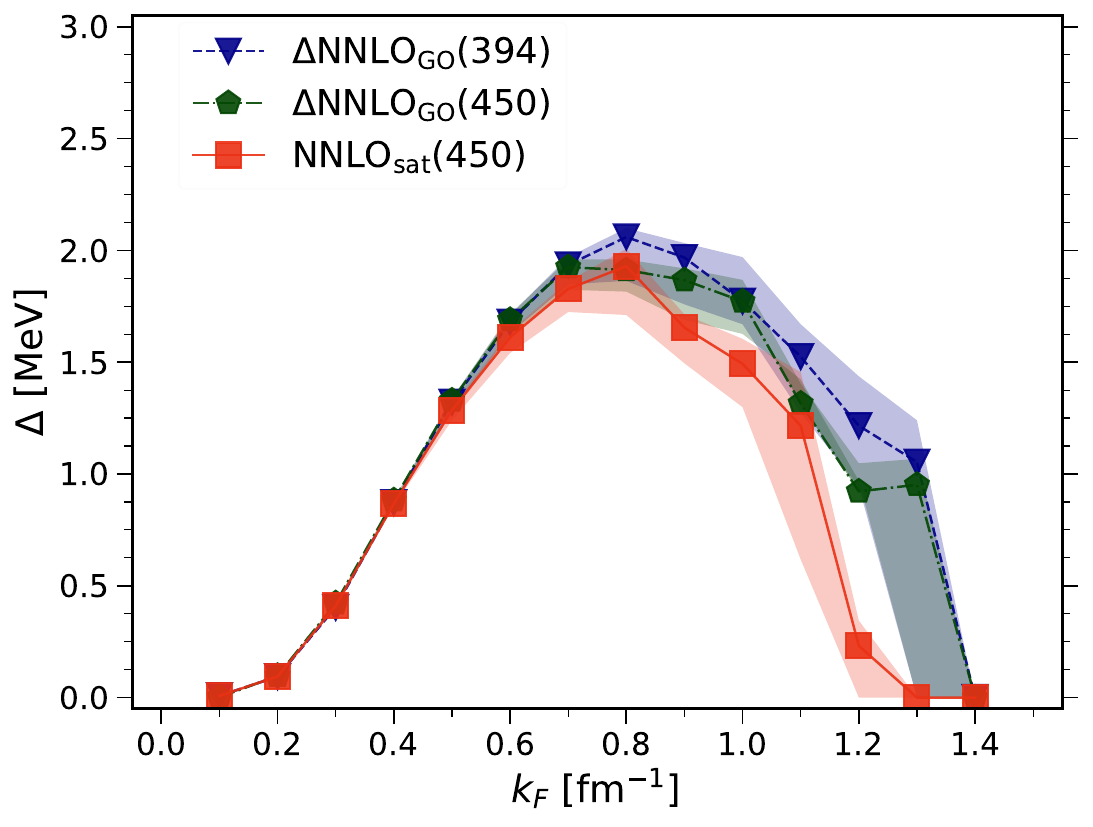}
    \caption{Pairing gaps as functions of the Fermi momentum.
    Calculations at the ADC(3) level are reported for the $\Delta \rm{NNLO}_{GO}(394)$ and $\Delta \rm{NNLO}_{GO}(450)$ potentials from Ref.~\cite{DeltaGo2020} and the $\rm{ NNLO_{sat}(450)}$ interaction.
    The shaded bands account for the uncertainties stemming both from the many-body truncation, estimated as the deviation between ADC(3) and ADC(ld,2) predictions, and the fit errors.
    Lines are a guide to the eye.
    }
    \label{fig:PairingGaps_Potentials}
\end{figure}

With the many-body truncations under control, we can now investigate the interaction dependence of the pairing gaps.
In Fig.~\ref{fig:PairingGaps_Potentials}, ADC(3) predictions are shown for three potentials at next-to-next-to-leading order in the chiral expansion (see legend) which reasonably describe the nuclear matter saturation properties~\cite{Marino2026Gorkov}.
These calculations are equipped with estimates of the theoretical errors, shown as shaded bands.
We account for 
both the uncertainty  associated with the fit
(as in Fig.~\ref{fig: Gaps_Truncations Adc}) and that related to the many-body truncation, which we assess as the deviation between ADC(3) and ADC(ld,2).

Remarkably, the ADC computations for all three chiral interactions are in excellent agreement at low Fermi momenta.
In this region, where the $S$-wave component of the nuclear interaction dominates, pairing gaps are strongly constrained by the $^{1}S_{0}$ phase shifts~\cite{Dean2003,Gezerlis2014Review}. This agreement persists beyond the mean-field approximation.
In fact, as long as a ``minimal'' set of higher-order correlations is incorporated, predictions up to at least $k_F$=0.8 fm$^{-1}$ are rather robust, and the many-body uncertainty is almost negligible. In light of Fig.~\ref{fig: Gaps_Truncations Adc}, we identify ladder diagrams as the crucial ingredient~\cite{Lombardo2001}.
The interaction dependence, instead, becomes increasingly pronounced above the position of the gap maximum, where details of the finite-range interaction, three-nucleon contributions, and many-body effects become significant.
This is consistent with larger theoretical uncertainties.

\begin{figure}[h!]
    \centering
    \includegraphics[width=\columnwidth]{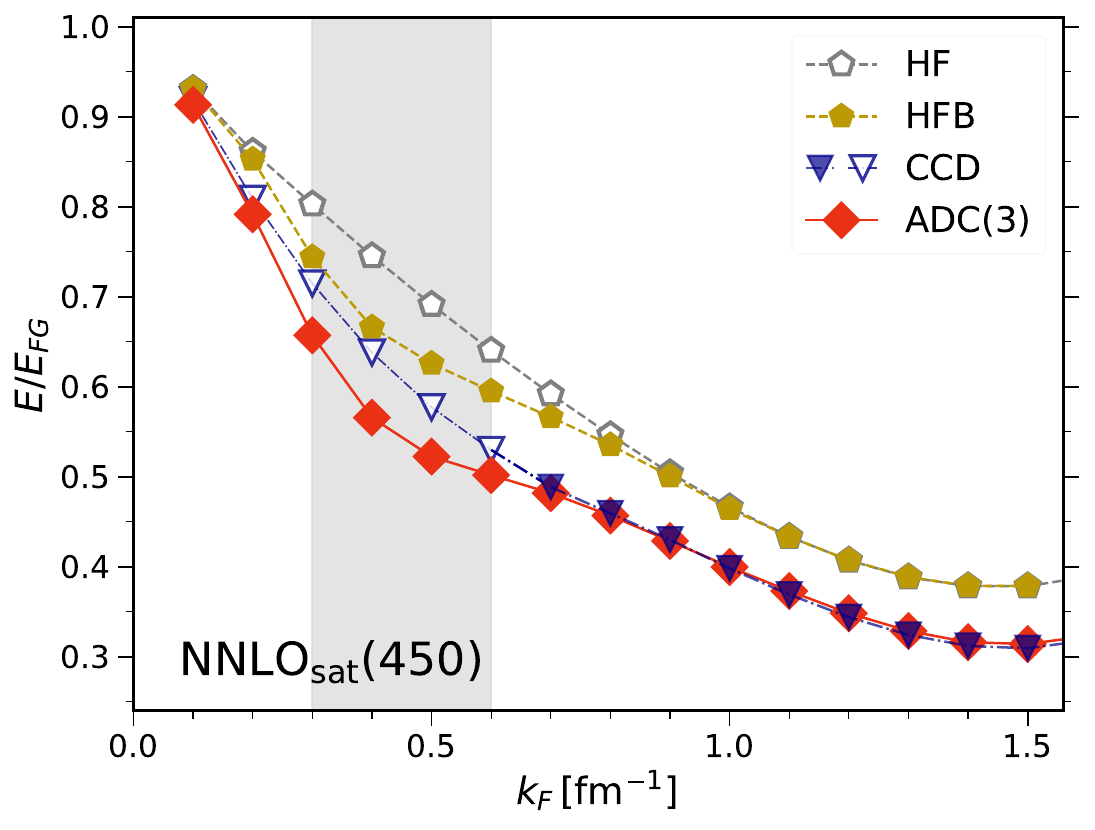}
    \caption{Energy per particle in units of the free Fermi gas energy $E_{FG} = 3/5 E_F$, where $E_F = (\hbar k_F)^{2}/(2m)$. 
    Results are obtained with the  $\rm{NNLO_{sat}(450)}$ interaction~\cite{NNLOsat}.
    Empty triangles denote extrapolations from the CCD energies (full triangles).
    The gray shaded area corresponds to the region where ADC(3) pairing gaps (shown in Fig.~\ref{fig:PairingGaps_Potentials}) satisfy $\Delta/E_F > 0.2$.
    }
    \label{fig: EOS Sat comparison methods}
\end{figure}

Finally, we elucidate the role of pairing correlations in Fig.~\ref{fig: EOS Sat comparison methods} focusing on the EOS.
The predictive power of many-body approaches that do not include pairing explicitly is expected to deteriorate in the superfluid regime, much as in open-shell nuclei. This can manifest itself through a reduced accuracy (in a variational sense) relative to methods that explicitly incorporate pairing correlations (see, e.g.,~\cite{Gezerlis2010,Kim:2023fwy,Urban2023}), or even through a complete breakdown of convergence~\cite{Cao2006,Sedrakian2019,Rios2020}.
We compare HFB and Gorkov-ADC(3) results to normal-state Hartree-Fock (HF) and coupled-cluster calculations at the doubles truncation level (CCD)~\cite{Hagen2014NucMat,LietzCompNucl}. The latter are based on PBCs.

Starting from the highest Fermi momenta shown, and moving toward lower $k_F$, the HF and HFB results remain nearly indistinguishable down to $k_F \approx 0.8 \, \mathrm{fm}^{-1}$. Below this value, however, the HFB solution becomes energetically favored, yielding lower energies and thus providing a superior variational description.
A similar trend emerges at the correlated level. CCD (full triangles) and ADC(3) agree closely around nuclear saturation density~\cite{Marino2024}. 
At Fermi momenta below 0.7 $\rm{fm}^{-1}$, however, the CCD equations fail to converge, hindering a direct comparison with ADC(3).

As a possible workaround, we extrapolate the CCD results using a novel density-continuation extension of the subspace-projected coupled-cluster algorithm~\cite{Ekstrom2019,Jiang2024Emulators}, described in the Supplemental material.
In this method, CCD solutions at selected Fermi momenta span a reduced subspace for continuing the normal-state coupled-cluster solution to other densities.
However, the emulator predictions (empty triangles) are unable to reproduce the subtle yet physically essential change in the concavity of the EOS observed at the ADC(3) level.
While previous studies have highlighted that subspace-projection techniques can capture crossovers between qualitatively different regimes~\cite{Frame:2017EigenvectorCont,Franzke:2023hdh}, our results indicate that superfluid solutions do not emerge from the normal-state CCD extrapolation considered here (see also~\cite{Sowinski:2022wdd}).

The discrepancy between ADC(3) and CCD is largest in the region $k_F \simeq 0.4$--$0.6\,\mathrm{fm}^{-1}$, where it reaches nearly $20\%$ of the total energy.
At very low densities, $k_F \sim 0.1$--$0.2\,\mathrm{fm}^{-1}$, the neutron gas becomes sufficiently dilute that interaction effects are strongly suppressed. 
Consequently, all many-body calculations approach the free Fermi gas energy $E_{FG}$.
The enhancement of the difference between the paired and normal solutions - HFB and HF at the mean-field level, ADC(3) and CCD at the correlated level - around $k_F \simeq 0.4$--$0.6\,\mathrm{fm}^{-1}$ is consistent with the density region in which the pairing gap is maximal (shaded band; see also Fig.~\ref{fig:PairingGaps_Potentials}), and thus constitutes an EOS-level signature of the superfluid phase. 

\runinhead{Conclusions and perspectives}
In this work, we have extended the \textit{ab initio} Gorkov-ADC method to superfluid neutron matter.
We have shown that superfluidity induces a distinct qualitative change in the EOS concavity that normal-state frameworks, such as normal-state coupled-cluster, cannot capture.
The pair formation mechanism has been elucidated by considering the energy- and momentum-resolved one-neutron spectral functions.
From these, we have evaluated the pairing gaps in the low-density regime, estimating the sensitivity to the many-body scheme by comparing different truncations in the ADC hierarchy.

Our findings demonstrate the role of dynamical self-energy effects in suppressing the pairing gaps in comparison to the mean-field~(HFB) prediction from the same Hamiltonian.
Once many-body correlations (in particular, ladder diagrams) are properly included, gaps are relatively well constrained up to about $k_F = 0.8$ fm$^{-1}$, showing a small residual method uncertainty and a remarkable consistency across different potentials.

Ultimately, Gorkov-ADC provides a single, robust framework for understanding neutron matter from low to high densities, giving valuable insights for modeling the neutron-star crust matter.
Our results pave the way for investigating the impact of pairing and many-body dynamical correlations on the electroweak processes that drive neutrino emission rates~\cite{Sobczyk2025,Steiner2009,Reddy1998,Shternin2018, Sedrakian2024_shortrange} and strongly affect the thermal evolution of neutron stars~\cite{Page_2004,SchmittShternin2017}.

\runinhead{Acknowledgements}
The authors thank Mehdi Drissi, Alexandros Gezerlis, Arnau Rios, Armen Sedrakian, and Michael Urban for discussions.
Support by the Deutsche Forschungsgemeinschaft (DFG, German Research Foundation) – Project-ID 279384907 – SFB 1245, and through the Cluster of Excellence “Precision Physics, Fundamental Interactions, and Structure of Matter” (PRISMA+ EXC 2118/1, Project ID 390831469) is acknowledged.
We acknowledge the CINECA award RespGF (HP10BQMECT) under the ISCRA initiative, for the availability of high-performance computing resources and support.
This work used the DiRAC Data Intensive service (DIaL3) at the University of Leicester, managed by the University of Leicester Research Computing Service on behalf of the STFC DiRAC HPC Facility (www.dirac.ac.uk). The DiRAC service at Leicester was funded by BEIS, UKRI and STFC capital funding and STFC operations grants. DiRAC is part of the UKRI Digital Research Infrastructure.
Part of the calculations were also performed at the supercomputer Mogon at Johannes Gutenberg Universit\"at Mainz.

\bibliographystyle{apsrev4-2}
\bibliography{bibliography.bib}


%
%

\clearpage
\onecolumngrid 
\begin{center}
    {\large\bfseries Supplemental Material}
\end{center}

\setcounter{equation}{0}
\renewcommand{\theequation}{S\arabic{equation}}

\setcounter{figure}{0}
\renewcommand{\thefigure}{S\arabic{figure}}

\setcounter{secnumdepth}{2} 
\setcounter{section}{0}
\renewcommand{\thesection}{S\arabic{section}}

\makeatletter
\setcounter{NAT@ctr}{0}
\makeatother

\begin{bibunit}[apsrev4-2]

\section{Single-particle basis and model-space convergence}
\label{sec: model-space convergence}

The calculations presented in this work are performed in a finite model space, employing a finite number of particles in a finite box (see~\cite{LietzCompNucl,Hagen2014NucMat,Marino2023,Marino2026Gorkov}). The single-particle basis states are defined as
\[
\ket{\alpha} = \ket{\mathbf{k}_\alpha, \sigma_\alpha},
\]
where $\sigma_\alpha = \pm 1/2$ denotes the spin projection (the isospin-projection is irrelevant for neutron matter.)
The allowed momenta are discretized according to~\cite{Marino2026Gorkov,CeperleyTABC}
\begin{align}
    \label{eq:momgrid}
    \mathbf{k} = \frac{1}{L} \big( 2\pi \mathbf{n} + \boldsymbol{\theta} \big),
\end{align}
where the box size is fixed by $L = (N/\rho)^{1/3}$ for a given density $\rho$ and neutron number $N$.
Throughout this work, we use $N=66$.
Here, $\mathbf{n}=(n_x,n_y,n_z)$ is a triplet of integers $n_i = 0, \pm1, \pm2, \dots$, and $\boldsymbol{\theta}$ denotes the twist-angle vector with components $0 \le \theta_i \le \pi$. Periodic boundary conditions (PBCs) are recovered for $\boldsymbol{\theta}=0$.
We use the special-point twisted-angle boundary conditions (sp-TABC) scheme~\cite{CeperleyTABC,Hagen2014NucMat,Marino2026Gorkov}, where the twist angles are chosen by minimizing the deviation between the Hartree--Fock energy of the finite system and the corresponding thermodynamic-limit value as a function of $\boldsymbol{\theta}$~\cite{Marino2026Gorkov,Hagen2014NucMat}. 

\begin{figure}[h!]
    \centering
    \includegraphics[width=0.6\columnwidth]{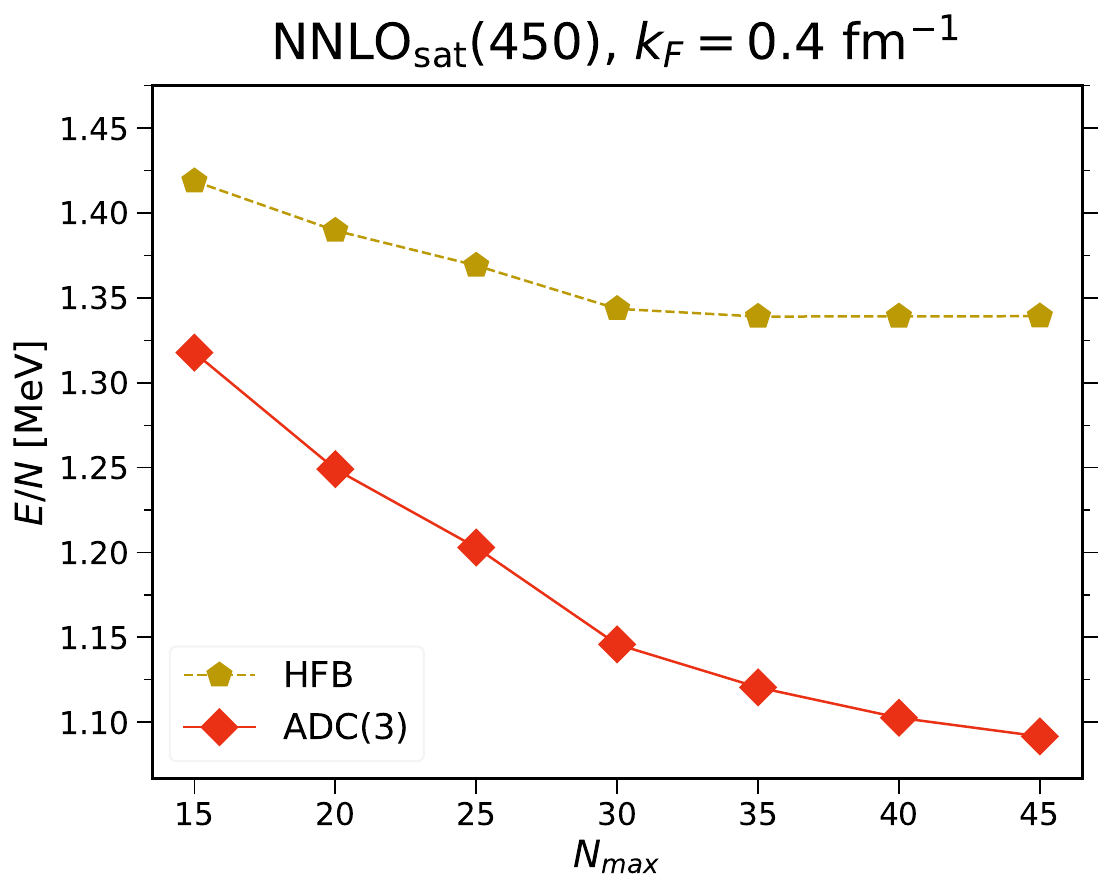}
    \caption{Energy per particle obtained with HFB and ADC(3) as a function of the model-space truncation parameter $N_{\max}$ for PNM at $k_F=0.4\,\mathrm{fm}^{-1}$. The $\rm{NNLO_{sat}(450)}$ interaction is employed.}
    \label{fig:Convergence_Pattern}
\end{figure}

We truncate the single-particle Hilbert space by including states that satisfy the condition~\cite{Marino2026Gorkov}
\begin{align}
    \abs{\mathbf{k}}^{2} \le \frac{4\pi^2}{L^2} N_{max}
\end{align}
The parameter $N_{max}$ is gradually increased until energies converge within a small tolerance. 
The dependence of the energy per neutron on the single-particle model-space dimension is displayed in Figure~\ref{fig:Convergence_Pattern}.
Calculations are performed using the relatively hard $\rm{NNLO_{sat}(450)}$ interaction~\cite{NNLOsat}.
We focus on the low-density case $k_F=0.4\,\mathrm{fm}^{-1}$, where convergence is expected to be particularly slow~\cite{Marino2026Gorkov}.
We report results for the HFB and ADC(3) truncations.
HFB energies are essentially converged already for $N_{\max}=35$, differing from the $N_{\max}=45$ results by a few keV per particle. ADC(3) exhibits a slightly slower convergence. However, at $N_{\max}=35$ the residual deviation from the largest model space remains below 20 keV, corresponding to less than 2\% of the total energy.
We therefore estimate the model-space truncation uncertainty for ADC(3) energies at $N_{\max}=35$ to be below 2\% over the density range considered. 
All ADC (at different levels of truncation) and CCD simulations in the main text are performed in a model space with $N_{\max}=35$. 
As HFB is considerably cheaper computationally, we employ $N_{\max}=40$ for this scheme.

\section{Density-continuation emulator for CCD}
To extend the CCD equation of state into the low-density region where the nonlinear amplitude equations cease to converge, for the first time we employ a density-continuation variant of the subspace-projected coupled-cluster method \cite{Ekstrom2019, Jiang2024Emulators} based on eigenvector continuation \cite{Frame:2017EigenvectorCont}. 
In the original formulation, the reduced space is spanned by coupled-cluster solutions obtained for different parameters of the nuclear interaction.
Here in the density continuation method, instead, the interaction is kept fixed and high-fidelity CCD calculations are performed at a small set of training Fermi momenta $k_F^{(i)}$ where the full CCD equations converge.

At each training point, the right and left CCD ground states are defined as~\cite{Hagen2014Review,Hagen2014NucMat}
\begin{align} \ket{\Psi_i} &= e^{T^{(i)}}\ket{\Phi_0}, \nonumber \\ \bra{\widetilde{\Psi}_i} &= \bra{\Phi_0} \left(1+\Lambda^{(i)}\right) e^{-T^{(i)}}, \end{align} 
where $\ket{\Phi_0}$ is the Hartree-Fock reference state and $T^{(i)}$ and $\Lambda^{(i)}$ are the converged CCD excitation and de-excitation amplitudes, respectively, at the training Fermi momentum $k_F^{(i)}$.

We keep the neutron number fixed to $N=66$, thus the box size $L$ varies as a function of the density~\cite{LietzCompNucl,Marino2026Gorkov} (Sec.~\ref{sec: model-space convergence}).
In particular,
\begin{equation} L = \left(\frac{N}{\rho}\right)^{1/3} = \frac{(3\pi^2 N)^{1/3}}{k_F}. \end{equation}
For the periodic boundary conditions used in the CCD calculations, the single-particle momenta can therefore be written as, cf. Eq~\eqref{eq:momgrid},
\begin{equation} 
\label{eq: states kF}
\mathbf{k}_{\mathbf{n}}(k_F) = \frac{2\pi\mathbf{n}}{L} = \frac{2\pi}{(3\pi^2 N)^{1/3}} k_F\,\mathbf{n}. \end{equation}

A potential concern is that, effectively, a different momentum basis is constructed at each training density. This differs from conventional eigenvector-continuation applications, where the single-particle basis is kept fixed for a given nucleus~\cite{Ekstrom2019} or at a given density~\cite{Jiang2024Emulators}. In the present case, however, all calculations employ the same neutron number \(N\) and model-space truncation \(N_{\max}\). The finite single-particle space can therefore be represented by the same set of integer momentum labels \(\mathbf{n}=(n_x,n_y,n_z)\) at every density. The physical momentum associated with a given label changes smoothly with \(k_F\) according to Eq.~\eqref{eq: states kF}, establishing a one-to-one correspondence between the single-particle states at different densities:
\begin{equation}
\mathbf{k}_{\mathbf{n}}(k'_F)=\frac{k'_F}{k_F^{(i)}}\mathbf{k}_{\mathbf{n}}\left(k_F^{(i)}\right).
\end{equation}
Hence, the CCD solutions obtained at different training densities can be represented in a common abstract configuration space defined by the same discrete particle--hole excitation labels. The underlying single-particle labels are the integer momentum triplets \(\mathbf n=(n_x,n_y,n_z)\). Changing \(k_F\) changes the physical momenta associated with these labels, but not the discrete configuration structure itself.

For a target Fermi momentum \(k'_F\), we denote by \(H(k'_F)\) the finite-volume representation of the same underlying nuclear Hamiltonian in this common discrete basis. The notation \(H(k'_F)\) does not imply that the nuclear interaction itself is density dependent. Rather, the matrix representation changes smoothly with \(k'_F\) because the box size and the physical momenta associated with the fixed integer labels change with density.

The Hamiltonian and norm kernels between two training CCD solutions \(i\) and \(j\) are formally
\begin{align}
H_{ij}(k'_F)&=\bra{\widetilde{\Psi}_i}H(k'_F)\ket{\Psi_j},\\ 
N_{ij} &= \braket{\widetilde{\Psi}_i } {  \Psi_j}. 
\end{align}
Using the coupled-cluster parametrization, these expressions can be evaluated as
\begin{align}
H_{ij}(k'_F)&=\bra{\Phi_0}\left(1+\Lambda^{(i)}\right)e^{X_{ij}}\overline{H}^{(j)}(k'_F)\ket{\Phi_0},\\
N_{ij}&=\bra{\Phi_0}\left(1+\Lambda^{(i)}\right)e^{X_{ij}}\ket{\Phi_0},
\end{align}
where $X_{ij}=T^{(j)}-T^{(i)},$ and 
$\overline{H}^{(j)}(k'_F)= e^{-T^{(j)}}H(k'_F)e^{T^{(j)}}$.
Thus, the cluster amplitudes entering the left and right training states remain those obtained from their respective training calculations, while the Hamiltonian entering the kernel is always the one evaluated at the target density \(k'_F\).

The subspace energy \(E(k'_F)\) and expansion coefficients \(\boldsymbol{c}\) are then obtained from the generalized non-Hermitian eigenvalue problem
\begin{equation}
H(k'_F)\,\boldsymbol{c}
=
E(k'_F)\,N\,\boldsymbol{c}.
\end{equation}

\begin{figure}
    \centering
    \includegraphics[width=0.6\columnwidth]{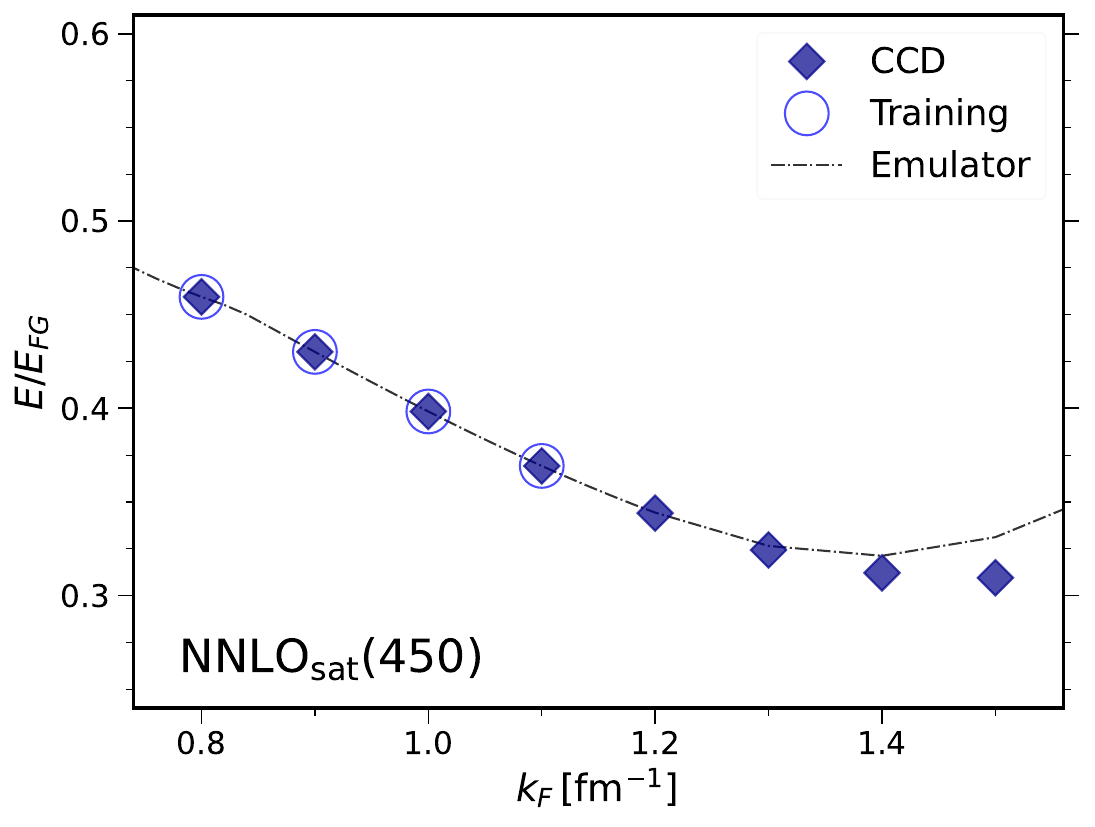}
    \caption{
    Energies in units of Fermi gas energy $E_{FG}$ for CCD (diamonds), compared to the predictions of the density-continuation emulator of the coupled-cluster amplitudes (dashed line).
    The training points are shown as circles.
    The $\rm{NNLO_{sat}}(450)$ potential is employed. The CCD calculations are based on periodic boundary conditions, and the model space is truncated at $N_{max}=35$.
    }
    \label{fig:Cfr_Emulator_Sat}
\end{figure}

This procedure bypasses the direct iterative solution of the CCD amplitude equations at the target density and therefore provides a reduced-space estimate even when the conventional CCD iterations become unstable or fail to converge. We find that only a small number of training states is sufficient to accurately interpolate the converged CCD results and to extrapolate the normal-state CCD branch over a broad range of densities, including densities well outside the training region. 
The performance of the density-continuation  algorithm is demonstrated in Fig.~\ref{fig:Cfr_Emulator_Sat}, where the predictions of the emulator (dashed line) are compared to full CCD computations (diamonds).
The emulator is rather accurate in interpolating the CCD results in the range of Fermi momenta where it has been trained on, with training points shown as circles.
It is also able to extrapolate at higher momenta, at least up to 1.4 $\rm{fm}^{-1}$, while the deviation with the full CCD calculations tends to increase at higher densities.


Since the reduced space is constructed exclusively from normal-state CCD solutions, however, the extrapolated energies must be interpreted as a continuation of the normal-state CCD equation of state. As shown in the main text, the density-continuation algorithm may not capture the emergence of pairing correlations.

Let us also comment on the convergence difficulties of the CCD equations.
The scaling relation Eq.\eqref{eq: states kF} suggests that, as also pointed out in Ref.~\cite{Marino2026Gorkov}, the excitation gap between particle and hole states becomes smaller as $k_F$ is reduced. 
These near-degeneracies generate small energy denominators~\cite{Barbieri2017} and, as in the intruder-state problems~\cite{Zheng,Kaldor}, may lead to anomalously large cluster amplitudes and ultimately divergent iterations.

\section{Quasiparticle expansion of the Gorkov equations}

Since in homogeneous matter the self-energy is diagonal in momentum, the Gorkov equation can be cast as an energy-dependent eigenvalue problem (see~\cite{Soma2011,Urban2026,Marino2026Gorkov}),
\begin{align}
    \label{eq: Gorkov fully dynamic}
    \begin{pmatrix}
        \xi_k + \widetilde{\Sigma}^{11}(k,\omega_q) & \Sigma^{12}(k,\omega_q) \\
        \Sigma^{21}(k,\omega_q) & - \xi_k - \widetilde{\Sigma}^{11}(k,-\omega_q)
    \end{pmatrix}
    \begin{pmatrix}
        \mathcal{U}_{k}^{q} \\
        \mathcal{V}_{k}^{q}
    \end{pmatrix} 
    = \hbar \omega_{q}
    \begin{pmatrix}
        \mathcal{U}_{k}^{q} \\
        \mathcal{V}_{k}^{q}
    \end{pmatrix}.
\end{align}
We consider in this section the general case where both the normal and anomalous self-energies are energy-dependent, 
\begin{subequations}
\begin{align}
    & \Sigma^{11}(k,\omega) = \Sigma^{11(\infty)}(k) + \widetilde{\Sigma}^{11}(k,\omega), \\
    & \Sigma^{12}(k,\omega) = \Sigma^{12(\infty)}(k) + \widetilde{\Sigma}^{12}(k,\omega),
    \\ 
    \label{eq: sigma 21}
    & \Sigma^{21}(k,\omega) = 
    \Sigma^{12(\infty)}(k)^{*} + \widetilde{\Sigma}^{21}(k,\omega).
\end{align}
\end{subequations}
Also, we have set $\xi_k = (\hbar k)^2/(2m) - \mu + \Sigma^{11(\infty)}$.
Due to the isotropy of homogeneous matter, quantities are functions only of $k = \abs{\mathbf{k}}$.
As we consider spin-saturated neutron matter, we can omit both the isospin and spin projections
\footnote{Under these assumptions, also, $\Sigma^{12(\infty)}$ is actually real. }.
In general, Eq.~\eqref{eq: Gorkov fully dynamic} admits multiple solutions for each momentum state, with eigenvalues $\omega_q$ and eigenvectors $\left( \mathcal{U}_{k}^{q}, \mathcal{V}_{k}^{q} \right)$.

Within a Green's function approach, one can define the Fermi surface in the thermodynamic limit as the solution to\footnote{The Gorkov formalism is based on a grand-canonical formulation, thus $\hbar\omega$ refers to energies measured from the chemical potential~\cite{Soma2011,Marino2026Gorkov}.} $g^{11}(k,\omega=0)^{-1}=0$ (see ~\cite{Luttinger1960, Fetter,Sensarma2007}), where $g^{11}$ is the normal component of the propagator $\mathbf{g}$~\cite{Marino2026Gorkov}.
This leads to the condition
\begin{align}
    \label{eq: chem pot condition}
    0 &= t_{k_F} - \mu + \Sigma^{11}(k_F, \omega=0) \\ 
    &= \xi_{k_F} + \widetilde{\Sigma}^{11}(k_F,\omega=0),
    \nonumber
\end{align}
which generalizes the HFB condition $\xi_{k_F}=0$~\cite{Sensarma2007}.

At the HFB mean-field level ($\widetilde{\Sigma}^{11} = \widetilde{\Sigma}^{12} = 0$), Eq.~\eqref{eq: Gorkov fully dynamic}
has the well-known solution~\cite{Fetter,Ring80a}
\begin{align}
    \label{eq: sol hfb pole}
    \hbar\omega_{k} = \pm \sqrt{ \xi_k^{2} + \abs{\SigAnom(k)}^{2} },
\end{align}
with a pair of eigenvalues ($\hbar\omega_{k}$) of opposite sign for each value of $k$.
Then, the pairing gap can be defined as the minimum of the dispersion band $\abs{\hbar \omega_k}$ as a function of $k$, 
\begin{align}
    \label{eq: gap definition}
    \Delta = \min_{k} \abs{ \hbar \omega_k }.
\end{align}
In the thermodynamic limit, $\xi_k$ vanishes as $k \to k_F$ (see above).
Thus, in HFB, $\Delta$ is equal to
\begin{align}
    \label{eq: gap as limit of sigma}
    \Delta = \lim_{k\to k_F} \abs{ \SigAnom(k) }.
\end{align}

The definition of the pairing gap in Eq.~\eqref{eq: gap definition} is well-motivated for both infinite and finite systems~\cite{Soma2012,Barranco2004} and can be naturally extended to beyond-mean-field theories by identifying it with the minimum energy of the fermionic quasiparticle branch~\cite{Haussmann2009,Magierski2009,Magierski2011, VanLoon2020}. 
In HFB, this branch is given by the Bogoliubov dispersion \(\omega_k\), Eq.~\eqref{eq: sol hfb pole}, whereas in our SCGF computations we identify it with the dominant quasiparticle poles of the Gorkov propagator.
We therefore associate, at each momentum, \(\omega_k\) with the solution carrying the largest spectral weight and define the gap from the minimum of \(|\hbar\omega_k|\)~\cite{Haussmann2009,Magierski2009}. 
Around the Fermi momentum, where this minimum occurs in the regime considered here, satellite poles tend to vanish and the spectral strength is strongly concentrated in a single pole close to the chemical potential (see main text), making the identification of the quasiparticles essentially unambiguous~\cite{Fetter}.

We now wish to find an explicit expression for the pairing gap. 
To this purpose, we perform a quasiparticle analysis of the Gorkov equations Eq.~\eqref{eq: Gorkov fully dynamic}.
We focus on the lowest-energy solution of Eq.~\eqref{eq: Gorkov fully dynamic}, which yields $\omega_k$ for $k$ close to the Fermi momentum.
This satisfies 
\begin{align}
    \label{eq: secular}
     \left[ \xi_k + \widetilde{\Sigma}^{11}(k,\omega_k) - \hbar \omega_k \right] \, \left[ - \xi_k - \widetilde{\Sigma}^{11}(k,-\omega_k) - \hbar \omega_k \right]
     - \Sigma^{12}(k,\omega_k) \Sigma^{21}(k,\omega_k) = 0.
\end{align}
Then, we linearize the dependence on $\omega_k$ in the lhs as follows. (A similar approach was adopted in Refs.~\cite{Barranco2004,Idini2012}.)
Assuming that $\omega_k$ is small enough, we can expand the dynamical self-energies around $\omega=0$ for simplicity.
Since the anomalous field is even in $\omega$~\cite{Urban2026,Soma2011},
\begin{align}
    \Sigma^{12}(k,\omega_k) = \Sigma^{12}(k,\omega=0) + \mathcal{O}(\omega_k^{2})
\end{align}
and similarly for $\Sigma^{21}$.
For the normal self-energy, instead,
\begin{align} 
    \widetilde{\Sigma}^{11}(k,\omega) = \widetilde{\Sigma}^{11}(k,0) + \omega_k \pdv{ \widetilde{\Sigma}^{11} }{\omega} \, (k,\omega=0). 
\end{align}
Thus, Eq.~\eqref{eq: secular} reads at leading order in $\omega_k$ as
\begin{align}
    \left( 1 - \frac{1}{\hbar} \pdv{ \widetilde{\Sigma}^{11} }{\omega} \right)^{2}
    (\hbar \omega_k)^{2} 
     - \left( \xi_k + \widetilde{\Sigma}^{11}(k,0) \right)^{2} - \abs{\Sigma^{12}(k,0)}^{2} = 0
\end{align}
where the derivative is evaluated for $(k, \omega=0)$.
Finally, we find a pair of solutions 
\begin{align}
    \label{eq: Hfb corrected}
    \hbar \omega_{k} = \pm \sqrt{ \widetilde{\xi}_{k}^{2} + \widetilde{\Delta}_{k}^{2} }
\end{align}
where
\begin{subequations}
    \begin{align}
        \widetilde{\xi}_{k} &= \mathcal{Z} \left( \xi_k + \widetilde{\Sigma}^{11}(k,\omega=0) \right), \\
        \widetilde{\Delta}_{k} &= \mathcal{Z} \left( \SigAnom(k) + \widetilde{\Sigma}^{12}(k,\omega=0) \right), \\
        \label{eq: quasiparticle weight}
        \mathcal{Z} &= \left( 1 - \frac{1}{\hbar} \pdv{ \widetilde{\Sigma}^{11} }{\omega} \, (k,\omega=0) \right)^{-1},
    \end{align}
\end{subequations}
where $\mathcal{Z}$ is the quasiparticle weight, and $\widetilde{\xi}_{k}$ and  $ \widetilde{\Delta}_{k}$ are the effective single-particle energy and effective pairing field, respectively.
Eq.~\eqref{eq: Hfb corrected} manifestly generalizes the standard HFB solution by incorporating dynamical self-energy corrections.
In particular, for $k \to k_F$, using Eq.~\eqref{eq: chem pot condition} one has $\widetilde{\xi}_{k} \to 0$ and Eq.~\eqref{eq: Hfb corrected} yields
\begin{align}
    \label{eq: Effective Gap}
    \hbar \omega_{k=k_F} &= \widetilde{\Delta}_{k=k_F} \\
    &=  \mathcal{Z} \, \Big( \SigAnom(k_F) + \widetilde{\Sigma}^{12}(k_F,0) \Big),
    \nonumber
\end{align}
which shows that the lowest-energy solution of the Gorkov equations (found at the Fermi momentum)---and thus, precisely the pairing gap as defined in Eq.~\eqref{eq: gap definition}---is quenched with respect to the corresponding expression in HFB by a factor $\mathcal{Z}<1$ (as $\pdv{\widetilde{\Sigma}^{11}}{\omega} < 0$).
This result, also known as the renormalized gap, was derived e.g.~in Refs.~\cite{Bozek2000,Lombardo2001,Bozek2003}.
(Notice that the authors of Ref.~\cite{Urban2026} have associated $\Delta$ with the value of the energy-dependent anomalous self-energy at the Fermi surface.)

Notice that the effect on the pairing gaps of dynamical dependencies in the anomalous sector are subleading with respect to including a fully dynamical normal self-energy. 
In particular, they amount to a constant shift $\widetilde{\Sigma}^{12}(k_F,0)$, see Eq.~\eqref{eq: Effective Gap}.
This result stems from the symmetry properties of the anomalous self-energy and  gives theoretical support to the scheme used in Refs.~\cite{Marino2024,Marino2026Gorkov} and in the present work, where the anomalous field is approximated at the static leading order. In this case, we obtain the result reported in the main text,
\begin{align}
    \Delta = \lim_{k \to k_F} \widetilde{\Delta}_{k} = \mathcal{Z}\, \SigAnom(k_F).
\end{align}


\putbib[supplBiblio]

\end{bibunit}

\end{document}